\documentclass{aa}
\begin{document}

   \thesaurus{8(09.07.1;10.08.1;12.04.1;12.04.2;11.17.1)} 
   \title{Updated Parameters for the Decaying Neutrino Theory and EURD
Observations of the Diffuse UV Background}
\titlerunning{Decaying Neutrino Theory}

   \author{D.W. Sciama
   \inst{1,}
   \inst{2,}
   \inst{3}}

   \institute{S.I.S.S.A., Strada Costiera 11, 34014, Trieste, Italy  (sciama@sissa.it)
    \and I.C.T.P., Trieste
 \and Department of Physics, Oxford University}
\date{}

   \maketitle

   \begin{abstract}

   Various  recent observational developments are here used to make a
more critical analysis of the parameter space of the decaying neutrino
theory for the ionisation of the interstellar medium. These developments
involve phenomena inside our Galaxy, outside the Galaxy but at essentially
zero red shift, and at large red shifts. This new analysis leads to a 
viable theory with a
decay lifetime of $2{\pm}1\times 10^{23}$ sec, a decay photon energy of
$13.7{\pm}0.1$ eV, and a mass for the decaying neutrino of
$27.4 {\pm}0.2$ eV. 

These parameters, when combined with some known astronomical quantities,
lead to predictions for the intensity, the wavelength and the width of the
decay line produced by neutrinos lying within one optical depth of the sun
$({\sim}{1\over2}\,{\rm pc})$. One finds an intensity in the line of 
$350^{+350}_{-117}$
photons cm$^{-2}$ sec $^{-1}$, a wavelength of $905\pm7\AA$, and a width
$\sim1\AA$. These predictions are relevant for the observations about to
be made by the EURD ultra-violet detector which is currently in Earth 
orbit on board the Spanish MINISAT 01 satellite.

   \end{abstract}

%

\section{Introduction}

A variety of recent observational developments, including Hipparcos data
and the probable discovery of the HeII Gunn-Peterson effect (Zheng et al
1998), make it desirable to update the parameters of the decaying neutrino
theory for the ionisation of hydrogen in warm opaque regions of the
interstellar medium (Sciama 1990a, 1995, 1997a). In particular we need to
prepare for the forthcoming observations of the extraterrestrial diffuse
background at wavelengths below $912\AA$ to be made by the EURD detector
(Bowyer et al 1995, Morales et al 1997, Bowyer, Edelstein \& Lampton 1997)
on board the Spanish MINISAT 01 satellite. This satellite was successfully
launched on April 21 1997, and at the time of writing the detector is
working well. One of its tasks is to search for a decay line emitted by
neutrinos within half a parsec of the sun.

The decaying neutrino theory makes a number of specific predictions for 
a variety of phenomena within individual galaxies,in intergalactic
space at red shifts in the range 0 to 5, and in the early universe at
red shifts between 5 and 1000. Two recent successes may be noted here.
The first is the observational verification of its prediction that the
density of free electrons in the interiors of warm opaque interstellar
clouds near the sun should be (a) substantial ($\sim 0.05$  cm$^{-3}$ ) and (b) the
same in each cloud (Spitzer \& Fitzpatrick 1993, Sciama 1997a).           
                                                                        
The second success concerns its rather precise predictions (Sciama 1997b) that
the Hubble constant should be $55\pm 0.5$ km sec$^{-1}$ Mpc$^{-1}$
and that the age of the
universe should be $12\pm  0.1$ Gyr. While the actual values of these
quantities remain controversial, it is noticeable that there has been a
general tendency recently for the higher estimates of both these
quantities to be significantly reduced. Moreover a number of these recent
estimates are in good agreement with our predictions, although the
observational uncertainties are still in the range $10$ to $20$ $\%$,
rather than our uncertainty of $1 \%$ .                         

The decaying neutrino theory is based on three parameters, namely, the
rest mass $m_\nu$ of the decaying neutrino, its radiative lifetime $\tau$,
and the monochromatic energy $E_\gamma$ of the decay photon in the rest
frame of its parent neutrino. For simplicity we follow popular (but
unproved)  particle physics models (such as the see--saw model (Yanagida
1978, Gell-Mann, Ramond \& Slansky 1979)) in which the secondary neutrino
in the decay has a much smaller mass than $m_\nu$. In that case
$E_\gamma={1\over 2}\,m_\nu$, and there are only two parameters to be
determined, namely, $\tau$ and $E_\gamma$.

In section 2 $\tau$ is determined from the observed ionisation in the
interstellar medium and from observational limits on the diffuse
extragalactic background at $1500\AA$. Its value is found to be
$2\pm1$x$10^{23}$ sec. In section 3 $E_\gamma$ is determined from upper limits on the
extragalactic hydrogen--ionising background at various redshifts to be $13.7\pm0.1$eV. Finally
section 4 discusses the implications of these results for the forthcoming
observations by EURD. In particular, with the help of additional
astronomical parameters, estimates
 are given for the intensity, the wavelength, and the width of the decay
line postulated to be emitted by neutrinos in the vicinity
$(\sim{1\over2}\,{\rm pc})$ of the sun. 


\section{The Decay Lifetime $\tau$}
\subsection{ The ionisation of the interstellar medium}
   An upper limit on $\tau$ will now be derived from recent estimates of
the maximum permitted density of dark matter near the sun and of the free
electron density $n_e$ which is here being attributed to the ionisation of
hydrogen by decay photons. 

In section $2.2$ a lower limit on $\tau$ will be determined from recent
observational upper limits on the extragalactic diffuse background at
$1500\,\AA$, which impose an upper limit on the flux of red shifted decay
photons emitted by the cosmological
 distribution of neutrinos.

Our derived upper and lower limits on $\tau$ are only just consistent with
one another, and so lead to a highly constrained value for this quantity. 

To derive the upper limit on $\tau$ consider a region near the sun whose
atomic hydrogen density makes it opaque to ionising decay photons. In
ionisation equilibrium one would have

 \begin{equation}
      {n_{\nu}\over{\tau}} = \alpha\,n^2_e,
   \end{equation}                    

\noindent
where $n_{\nu}$ is the local number density of neutrinos and $\alpha$ is
the recombination coefficient (excluding recombinations directly to the
ground state). The value of $n_e$ which the theory attributes to
ionisation by decay photons has been recently rediscussed by Sciama (1997a).
Using observations of dispersion measures for nearby pulsars with known
radio parallaxes (Gwinn et al 1986, Bailes et al 1990), and HST
observations of the absorption spectrum of the halo star HD93521 
(Spitzer \& Fitzpatrick
1993) (which are somewhat less straightforward to interpret), 
the result obtained was $n_e=0.05\pm0.01\rm {cm}^{-3}$ (Reynolds 1990,
Sciama 1990 b, 1997a).

The value of $\alpha$ depends on the temperature T of the gas. For the
relevant regions of the interstellar medium with the better determined
values of $n_e$ one has $T<2$x$10^4\,K$ (Reynolds 1985). Hence
$\alpha>1.3\times10^{-13}\rm{cm}^3\,\rm{sec}^{-1}$ (Ferland et
 al 1992).

Finally, to derive an upper limit on $\tau$ an upper limit on the number
density $n_\nu$ must be established. This can be obtained from an upper
limit on the local mass density of neutrinos $\rho_{\nu}$, since earlier
discussions of the decaying neutrino theory have already provided a
sufficiently accurate value for $m_\nu$, namely $\sim28 \rm{eV}$ (Sciama
1990 a, 1995). An upper limit on $\rho_\nu$ near the sun was recently
derived by Sciama (1997 b) in connexion with values for the largest
permitted flattening of the neutrino halo and for the rotational velocity
of the Galaxy at the sun's position. By taking into account various
estimates for column densities of material near the sun an upper limit for
$\rho_{\nu}$ of $0.03$ $M_{\odot}$pc$^{-3}$ was obtained. This upper limit
would be associated with the largest permitted flattening of the dark halo
(Dehnen \& Binney 1997), corresponding to an axial ratio of 0.2 (shape E8).
It is reassuring that some other galaxies do seem to exhibit 
a similar flattening (Sackett et al 1994, Olling 1996, Becquaert \&
Combes 1997).

The local mass density $\rho_{\nu}$ can also be determined by comparing
dynamical estimates of the total density $\rho_o$ near the sun (the Oort
limit) with the densities of known stars and gas. The value of $\rho_o$
has been controversial; an estimate will be used here which is based on
recent Hipparcos observations of F stars, and so may lead to a more
reliable result. This result is $\rho_o=0.11\pm0.01 M_\odot$ pc$^{-3}$
(Pham 1996, 1997). However, a lower value for $\rho_o$ has recently been
obtained from Hipparcos data by Cr\'ez\'e et al (1998), namely
$0.076\pm0.015 M_\odot$ pc$^{-3}$. This result seems rather low when
compared to the density of known matter (see below) and the disagreement
remains to be clarified. Here we provisionally adopt Pham's value. 

The contribution of known stars and gas to $\rho_o$ is itself somewhat
uncertain. Often quoted values for each are $0.04 M_\odot$ pc$^{-3}$ (e.g. 
Bienaym\'e, Robin \& Cr\'ez\'e 1987, Cr\'ez\'e et al 1998). The stellar
contribution may have to be increased to allow for the existence of faint
stars, but the HST deep survey suggests that this additional contribution
may be small (e.g. Gould et al 1996). If this is correct, one again
obtains an upper limit for $\rho_{\nu}$ of $0.03 M_\odot$ pc$^{-3}$. This
upper limit was previously derived by Bienaym\'e et al (1987). If now
$m_{\nu}=27.4\,{\rm eV}$, as derived in section 2.2, there follows an
upper limit for $n_\nu$ near the sun of $4.16\times 10^7{\rm cm}^{-3}$.

Our limits on the values of $n_\nu$, $\alpha$ and 
$n_e(\geq 0.04 {\rm cm}^{-3})$
now lead, in conjunction with (1), to the conclusion that

   \begin{equation}
	\tau \le 2\times 10^{23}\; {\rm sec}.
   \end{equation}
This result can be shown to be compatible with our interpretation of
Reynolds' (1984) global H$\alpha$ data for the Galaxy (Sciama 1997b).

\subsection{The extragalactic background at 1500 $\AA$}
Some of the earliest lower limits on $\tau$ (for a decaying neutrino not
then related to the ionisation of the interstellar medium) were based on
observational estimates of the cosmic background in the far ultra-violet,
which was compared with the red shifted decay flux produced by the
cosmological distribution of neutrinos (Stecker 1980, Kimble, Bowyer \&
Jakobsen 1981). Estimates of this background are still controversial
(compare Bowyer 1991 with Henry 1991). Recent contributions to the
discussion have been made by Henry \& Murthy (1993), Witt \& Petersohn
(1994) and Witt, Friedmann \& Sasseen (1997). We adopt from their
discussions an upper limit of about 300 photons cm$^{-2}$ sec$^{-1}$
ster$^{-1}\AA^{-1}$ (continuum units or CU) at $1500\AA$. From
 this value one must subtract the contribution due to galaxies, which has
been evaluated by Armand, Milliard and Deharveng (1994) as about $40-130
\,CU$, leaving an upper limit of about $200 \,CU$ available for the decay
flux.

Theoretical aspects of this flux have been re-discussed by Sciama (1991),
Overduin, Wesson \& Bowyer (1993), Dodelson \& Jubas (1994) and most
recently by Overduin \& Wesson (1997). These calculations show that a decay
flux of $200\, CU$ at $1500\,\AA$ corresponds to a lifetime $\tau$ of
$2\times 10^{23}$ sec. Hence

\begin{equation}
\tau\ge 2\times 10^{23}\; {\rm sec}.
\end{equation}

 Our overall conclusion from this discussion is that, if the decaying
neutrino theory for the ionisation of the interstellar medium is correct,
the decay lifetime $\tau$ is determined as 

\begin{equation} \tau \sim 2\times 10^{23} \;{\rm sec}.  \end{equation} The
uncertainty in this estimate is difficult to pin down. A reasonable guess
would be $50 \%$ at most. Hence we adopt

\begin{equation}
\tau = 2\pm1\times 10^{23}\; {\rm sec}.
\end{equation}
We note, however, that if the Cr\'ez\'e et al (1998) value for $\rho_o$ 
is correct no solution for $\tau$ is possible, and the 
decaying neutrino theory would be ruled out.

\section{The Photon Energy $E_{\gamma}$}

\subsection{The $H_\alpha$ flux from intergalactic HI clouds}

Existing attempts to observe $H_\alpha$ radiation from opaque
intergalactic HI clouds lead to an upper limit on the 
hydro\-gen--ionising photon
flux F incident on the clouds arising from the cosmological distribution
of neutrinos. This upper limit on $F$ then leads
to an upper limit on $E_\gamma$ because of the role of 
the red shift in reducing the energy of a decay photon 
to below $13.6$ eV. Write

\begin{equation}
E_\gamma=13.6+\epsilon\;{\rm eV}.
\end{equation}
Then, if ${\epsilon\over{13.6}}\ll1$, F at zero red shift $F(0)$ is given by

\begin{equation}
F(0)={{n_{\nu}(0)}\over{\tau}}{c\over{H_o}}{\epsilon\over{13.6}},
\end{equation} 
where $n_{\nu}(0)$ is the standard cosmological number
density of neutrinos at $z=0$, (namely $3/11$ of the number density of
photons in the cosmic microwave background), $c$ is the velocity of light,
and $H_o$ is the present value of the Hubble constant. In the decaying
neutrino theory $H_o=55\pm0.5$ km sec$^{-1}$Mpc$^{-1}$ (Sciama 1997c). 

The $H_\alpha$ upper limit observations of Vogel et al (1995) and of
Donahue, Aldering \& Stocke (1995) imply that $F(0)\le 10^5$ photons
cm$^{-2}$ sec$^{-1}$. Since $n_{\nu}(0)=112.6\pm0.5\,\rm {cm}^{-3}$ (Sciama
1997c) it follows that

\begin{equation}
{\epsilon\over{\tau}}\le 7\times10^{-25}\;{\rm eV}\;{\rm sec}^{-1}.
\end{equation}
Combining this inequality with our previous value for $\tau$, one obtains

\begin{equation}
\epsilon\le 0.2\;{\rm eV},
\end{equation}
so that indeed ${\epsilon\over{13.6}}\ll1$. It follows that

\begin{equation}
E_\gamma \le 13.8\;{\rm eV}.
\end{equation}
 We need not be too surprised that $E_\gamma$ is required to
be so close to the ionisation potential of hydrogen, since cosmological
considerations alone had previously suggested that $m_{\nu}\sim28$ev, so
that ${1\over2}m_{\nu}\sim14$ eV, which is within $3 \%$  of $13.6$ eV.

The upper limit on $F(0)$ is based on the assumption that the
intergalactic clouds are so opaque that essentially every incident
ionising photon actually produces a free electron which then recombines
with a proton. However, the clouds concerned may contain holes in their
HI distribution which would then permit a larger ionising flux to be
compatible with the observed upper limits on the $H\alpha$ flux
(Bland-Hawthorn 1996). The covering factor of the clouds concerned is in
fact unknown. It is therefore desirable to establish an independent upper
limit on $F(0)$. Such a limit can be obtained from constraints on the
hydrogen-ionising flux at red shifts in the range $2$ to $4.5$. These
constraints can be obtained in two independent ways: (i) from the 
proximity effect in Lyman $\alpha$ clouds at these red shifts 
(Bajtlik, Duncan
\& Ostriker 1988) and (ii) from the recently observed absorption 
by He II at $304 \,
\AA$ in the spectra of high red shift QSOs. 

\subsection{The proximity effect in Lyman $\alpha$ clouds}
The proximity effect is the
reduction in the number of Lyman $\alpha$ clouds near a QSO (Murdoch et al
1986). This effect can be used to derive an upper limit on the ionising
decay flux F(z) at a red shift z in the range $2\,\rm{to}\, 4.5$. The
relation $F(z)=(1+z)^{3\over2} F(0)$ (Sciama 1990a) then leads to an upper
limit on $F(0)$. Absorption of decay photons by intergalactic clouds does
not have to be taken into account in deriving this relation even at $z\sim 2$ 
or greater, where in principle the absorption would be much larger
than at $z\sim 0$ (eg. Haardt \& Madau 1996), because with our small
value for $\epsilon$ the red shift reduces the energy of a decay photon to below
$13.6$ eV in a shorter distance than the absorption mean free path
even at large z. 

The proximity effect was first used by Bajtlik, Duncan \& Ostriker (1988)
to derive the background ionisation rate $\Gamma$ per H atom for
$1.7<z<3.8$ by attributing this effect to the additional and known direct
ionising influence of the QSO on nearby clouds. They obtained
$\Gamma\sim3\times10^{-12}\,\rm {sec}^{-1}$, with an uncertainty of a factor 3
either way. They also found that, within their uncertainty, $\Gamma$ did
not vary significantly with z. Of the many later attempts to determine
$\Gamma$ and its possible variation with z, the most recent are due to
Giallongo et al (1996), Lu et al (1996), Savaglio et al (1997) and Cooke,
Epsey \& Carswell (1997). Despite the considerable remaining uncertainties
there is a consensus that $\Gamma\sim2\times10^{-12}\,\rm{sec}^{ -1}$, with no
significant z variation out to $z=4.5$. 

There has been much discussion in the literature as to whether the
population of QSOs alone can provide a sufficient ionising flux to
account for $\Gamma$. According to Haardt \& Madau (1996)
$\Gamma_{\rm QSO}\sim10^{-12}\,\rm {sec}^{-1}$ at $z \sim 2$, but only
$2\times10^{-13}\,\rm {sec}^{-1}\rm {at}\,z\sim 4.5$. 
The discrepancy of a factor
$\sim 10$ at $z\sim 4.5$ appears to be significant, even if the one at
$z\sim 2$ lies within the uncertainties. 

All the estimates of $\Gamma$ from the proximity effect which have so far
been made have been based on the simple assumption that the influence of
the QSO on the nearby clouds is entirely due to the additional direct
ionisation which it produces. However,
 as Miralda-Escud\'e \& Rees (1994) (MR) pointed out, one should 
 also allow for the additional heat input due to the radiation from the QSO. 
 The resulting expansion of the clouds reduces their electron density, 
 and the increase of temperature reduces the recombination coefficient,
 so that the recombination rate of the clouds is reduced. The net
additional ionisation produced by the QSO is thus increased, and so the
implied value of the background ionisation rate is also increased. MR
further pointed out that, since the main contribution to the additional
heating comes from the ionisation of He II in the clouds, the effect is
larger where the QSO (but not the general UV background) is capable of
ionising He II. This remark has become particularly pertinent now, because
it seems that ionisation breakthrough for He II in the intergalactic
medium may not have occured until $z\sim 3$ (Songaila \& Cowie 1996, Hogan et al
1997, Reimers et al 1997, Boksenberg 1998, Songaila 1998), a situation which was
theoretically anticipated by MR and by Madau \& Meiksin (1994). 

While the numerical value of the MR effect is model dependent (preliminary
attempts to estimate it having been made by MR themselves) it is clear
that it has two consequences for our discussion. It leads to an increase
in the (small) discrepancy between $ \Gamma$ and $\Gamma_{\rm QSO}$ at $z\sim
2$, and to an increase of $\Gamma$ with z, in contrast to the rapid
decrease in $\Gamma_{\rm QSO}$. 

It has often been suggested that any discrepancy between $\Gamma$ and
$\Gamma_{\rm QSO}$ could be resolved by appealing to ionising radiation
emitted by hot stars in galaxies (eg. Miralda-Escud\'e and Ostriker 1990).
Recent discussions of this possibility have been given by Giroux \&
Shapiro (1996) and by Madau \& Shull (1996).  Metal-enrichment arguments
imply that if $\sim 25 \%$  of the Lyman continuum photons emitted by
hot stars escape from their parent galaxies, then these
 photons would be responsible for an ionisation rate $\sim10^{-12}\,\rm
{sec}^{-1}$ per H atom at $z\sim 3$. The actual escape fraction is not
known at high z, but at $z\sim 0$ it is less than $1 \%$ (Deharveng et
al 1997). This strong constraint was deduced by relating the $H_\alpha$
luminosity density of star-forming galaxies in the local universe (Gallego
et al 1995) to the $H_\alpha$ observations of Vogel et al (1995) and
Donahue et al (1995) which, as already mentioned, lead to an upper limit
on the hydrogen-ionising flux at $z\sim 0$. Although the escape fraction
is not known at high z, there exists evidence for considerable dust
extinction in galaxies at these redshifts (Meurer 1997, Cimatti et al
1997). Absorption by atomic hydrogen in these galaxies may also be
important.                                                            
                                                                      
This argument has recently been much streng\-thened by the
observations of Spinrad et al (1998) which failed to detect any Lyman
continuum radiation escaping from $z>3$ Lyman-limit galaxies. According to
these authors it is implausible that ionising radiation from young
galaxies can replace QSO ionisation at $z>4$.                                                                                      
 
Another much-discussed possibility is that radiation from stars at red
shifts much greater than 5 (the so-called Population III stars)  might
make an appreciable contribution to $\Gamma$ and be responsible for the
reionisation of the universe at $z>5$. Recent discussions of this
possibility have been given by Haiman \& Loeb (1997), Gnedin \& Ostriker
(1997) and Gnedin (1998). An important aspect of this proposal concerns
the level of metallicity which would result from the required stellar
activity. It is not clear whether this level agrees with observation,
especially in view of the low metallicity recently derived by Songaila
(1997) for the Lyman $\alpha$ clouds. We cannot go into this intricate question
here,and since we are seeking an upper limit on the intergalactic flux of
decay photons, it will now be assumed that most of the missing ionising
photons at $z\sim 2\; \rm {to}\; 4.5$ are produced by the cosmological
distribution of decaying neutrinos and not by hot stars.

This assumption would enable us to understand why the usual interpretation
of the proximity effect leads to a value of $\Gamma$ which is
approximately independent of z. If, at $z\sim 2$, $\Gamma_{\nu}$ is of the
same general order as $\Gamma_{\rm QSO}$ then, as z increases, 
$\Gamma_{\nu}$ would increase as ${(1+z)}^{3\over 2}$ while 
$\Gamma_{\rm QSO}$ decreases, leaving $\Gamma$ approximately constant. 

It would seem that a rough upper limit for $\Gamma_\nu$ can be 
derived by setting
$\Gamma_\nu\sim 2\,\Gamma_{\rm QSO}$ at $z=2$. Then, since at this red shift 
$\Gamma_{\rm QSO}\sim\,10^{-12}\,{\rm sec} ^{-1}$ according to Haardt \& Madau (1996), $\Gamma_\nu$ would $\sim2\times 10^{-12}\,{\rm sec}^{-1}$ and $\Gamma\sim 3\times 10^{- 12}\,{\rm sec} ^{-1}$. This
excess of $\Gamma$ over the proximity effect value of $\sim2\times 10^{-12}\,
{\rm sec}^{-1}$ could be attributed to the MR effect. Indeed, since in this
picture $\Gamma_\nu$ is the dominant contributor to $\Gamma$, the spectrum of the
QSO would differ appreciably from that of the background, and the MR
effect would then be greater than in the pure QSO case (Rees 1990, Sciama
1995). Finally, it should be noted that, with our adopted values, $\Gamma$
would increase by a factor $1.7$ between $z=2$ and $z=4.5$. This increase
could be consistent with the dependence of the MR effect on z. 

Our upper limit on $\Gamma_\nu$ at $z=2$ implies an upper limit on $\Gamma_\nu$ at $z=0$ of
$4\times10^{-13}\,{\rm sec} ^{-1}$. Since the decay photons have an energy close
to the Lyman limit this ionisation rate converts to a photon flux by using
the photoionisation cross-section
 at this limit, which is $6\times10^{-18}\,{\rm cm}^2$. Hence one obtains
$F(0)\le 7\times 10^4{\rm cm}^{-2}{\rm sec}^{-1}$, which is not essentially
different from the upper limit $\sim 10^5 {\rm cm}^{-2}{\rm sec} ^{-1}$
derived by Vogel et al (1995) and by Donahue, Aldering \& Stocke (1995),
the precise value of which in fact depends on the uncertain shapes of the
intergalactic clouds which they observed.

\subsection{{\rm HeII} absorption at $z\sim 3$}

We now show that the recently derived Gunn-Peterson optical depth in
HeII at $z\sim3$,  $\tau_{{\rm GP,HeII}}$ (Zheng et al 1998),
in conjunction with the known upper
limit on $\tau_{{\rm GP,HI}}$, leads to a lower limit on 
$\Gamma$ at $z\sim 3$. It turns out that this
lower limit is close to the upper limit estimated in Sec. 3.2.               
                                                                        
There now exist a number of observations of HeII absorption at $z\sim 2$ to $3$ 
in the spectra of QSOs (Jakobsen et al 1994, Tytler et al 1995, Jakobsen 
1996, Davidsen, Kriss \& Zheng 1996, Hogan, Anderson \& Rogers 1997, Reimers
et al 1997). There has been considerable discussion in these and other
papers (Madau \& Meiksin 1994, Fardal, Giroux \& Shull 1998) as to whether
this absorption is entirely due to the HeII in Lyman $\alpha$ clouds,or
whether part of it must be attributed to the Gunn-Peterson effect
arising in an essentially diffuse intergalactic medium. We here follow
the calculations of Zheng, Davidsen \& Kriss (1998), which lead to a
definite value of $\tau_{{\rm GP,HeII}}=1$ for this effect at $z=3$. 
This would imply that $n_{\rm HeII}(3)= 4\times 10^{-10}$ cm$^{-3}$. To
see whether this result is reasonable we derive from it the implied value
of the total diffuse intergalactic gas density $n(3)$ at $z=3$,
using estimates
for the HeII-ionising flux due to QSO radiation filtered through the
absorbing medium of Lyman $\alpha$ clouds and Lyman 
limit systems (Haardt \& Madau
1996, Fardal, Giroux \& Shull 1998), and the value 0.08 for the He/H number
ratio. Using $\Gamma_{\rm HeII}=6 \times 10^{-15}$ sec$^{-1}$ one obtains 
$n(3)= 4.7\times 10^{-6}$ cm$^{-3}$. Comparing this with the
higher of the two competing values for the total baryon density
$n_b(3)$, based on measurements of the deuterium abundance and the theory of
big bang nucleosynthesis (Schramm \& Turner 1998), one finds that
$n(3)/n_b(3)=0.36$.

This result is compatible with the somewhat model-dependent estimate of
$\Omega_{\rm Ly\alpha}(3)$
made by Giallongo, Fonta\-na \& Madau (1997). These authors found that at 
$z\sim 3$ about half of $n_b$ could be attributed to gas in Ly$\alpha$
clouds, leaving about
half for the IGM (since the contribution from galaxies can here be
neglected (Persic \& Salucci 1992)). Given the uncertainties, this fraction
of 1/2 is compatible with our derived value of 0.36.

The next step is to use the $1 \sigma$ upper limit of 0.04 for 
$\tau_{{\rm GP,HI}}(3)$ 
(Giallongo, Cristiani \& Trevese 1992). Since

\begin{equation}
\frac{\tau_{{\rm GP,HeII}}}{\tau_{{\rm GP,HeI}}}=0.1\, \frac{\Gamma_{\rm HI}}
{\Gamma_{\rm HeII}}
\end{equation}
it follows that
\begin{equation}
\Gamma_{\rm HI}\ge 250 \,\Gamma_{\rm HeII}, 
\end{equation}
so that $\Gamma_{\rm HI}\ge 1.5 \times 10^{-12}\; {\rm sec}^{-1}$.
This lower limit is compatible with the upper limit for 
$\Gamma_{\rm HI}$ ($\sim 3\times 10^{-12}$ sec$^{-1}$) proposed in
Sec. 3.2.

This comparison is not strictly self-consistent because, by choosing a
value of $\Gamma_{\rm HI}$ greater than that due to QSOs, 
a disturbance has been
introduced into the calculation of the opacity of the universe,since the
ionisation state of the absorbers would be affected. This disturbance
would be reduced if in fact $\Gamma_{\rm HeII}$ from QSOs had 
to be increased by a factor
of the same order as the ratio $\Gamma_{\rm HI}/\Gamma_{\rm QSO} \sim 3$,
since then $\Gamma_{\rm HI}/\Gamma_{\rm HeII}$ would not be much
altered.

An increase in the HeII - ionising power of QSOs over that arising
from the usual power-law spectrum has already been proposed by Sciama
(1994), who needed to ensure that an increase in $\Gamma_{\rm HI}$ due to decay
photons would not drive the universe to become completely opaque at the
HeII edge. This
 proposal was based on existing observational hints that many QSOs
possess a soft x-ray excess in their spectra. This excess was attributed
to a Guilbert-Rees (1988) thermal bump with $T\sim50$ eV, resulting from
the reprocessing of harder x-rays from the central regions of QSOs by
optically thick cold material. Since 1994 further observational evidence
has accumulated for the prevalence of a soft x-ray excess in the spectra
of QSOs. This evidence has been reviewed by Gondhalekar, Rouillon-Foley
\& Kellett (1996). It is also noteworthy that a $50$ eV bump would fit
nicely in the gap (due to galactic absorption) in the composite spectrum
shown in fig.6 of Laor et al (1997). On the theoretical side it has been
found recently that a slim accretion disk
 around a black hole at the centre of a QSO would produce a soft x-ray
excess without any Guilbert-Rees reprocessing (Shimura \& Takahara 1995,
Szuszkiewicz 1996, Szuszkiewicz, Malkan \& Abramowicz 1996).

These arguments have recently been strengthened by the considerations of
Korista, Ferland \& Baldwin (1997) who pointed out that a QSO emission
spectrum without a bump at 50 ev would not account (via excitation
effects) for the observed strengths of the HeII emission lines in QSO
spectra. These authors suggested that either the QSOs have a suitably
complicated geometry, or that their emission spectrum contains a
significant bump in the vicinity of the HeII ionisation edge at 54.4 ev.

A rough estimate for the resulting increase in $\Gamma_{\rm HeII}$ 
 led to a factor $\sim3.6$ (Sciama 1994). If the actual factor were
closer to 2 the absorption analysis would still not be much
changed, while the lower limit on $\Gamma_{\rm HI}$ would be 
increased to $3\times 10^{-12}$ sec$^{-1}$ which is
the same as our proposed approximate upper limit. A self-consistent
solution is thus possible. For this solution one would have
$n(3)/n_b(3)=0.5$, which is in good agreement with the estimate implied by 
the calculations of Giallongo, Fontana \& Madau (1997).

Our proposed introduction of an appreciable flux of decay photons at high
$z$ also has implications for the abundance of HeI at these red shifts.
It was argued by Miralda-Escud\'e \& Ostriker (1992) and by Reimers et al
(1993) that the low values of $N_{\rm HeI }$ observed in Lyman limit systems
and Lyman $\alpha$ clouds are incompatible with the decaying neutrino
theory. However, Sciama (1994) showed that if QSOs possess a soft x-ray
bump in their spectra, the resulting high ionisation of HeII in the
various cloud systems would sufficiently lower their abundance of HeI.
A further reduction in $N_{\rm HeI}$ could arise from hot stars in galaxies,
since the escape fraction of HeI-ionising photons would be expected to
exceed that of HI-ionising photons. Accordingly the existence of an
appreciable flux of decay photons at high $z$ is not incompatible with the
observed values and upper limits on $N_{\rm HeI}$. 

Our conclusion from all these considerations is that it is unlikely that
F(0) can be increased by a substantial factor, say $\sim2$, over the
upper limit derived from the $H\alpha$ measurements of intergalactic
clouds, by appealing to a small HI covering factor in these clouds.
Accordingly the constraint $E_{\gamma}\le 13.8$ eV still holds good, so that

\begin{equation}
E_\gamma=13.7\pm 0.1\,\rm eV
\end{equation}

\begin{equation}
m_\nu=27.4\pm 0.2\,\rm eV,
\end{equation}
and, from section 2

\begin{equation}
\tau=2\pm 1\times10^{23}\,\rm sec.
\end{equation}

These are our updated parameters for the decaying neutrino theory. It
should be noted that these parameters lead to precise values for the
Hubble constant $H_0 \;(55\pm0.5\,\rm {km\,sec}^{-1}\rm {Mpc}^{-1})$ and
the age of the universe $(12\pm 0.1$ Gyr) (Sciama 1997c), if the
cosmological constant is zero.

\section{Predictions for the EURD Observations}

We now predict the intensity, the wavelength and the width of the decay
line due to neutrinos near the sun. The uncertainties which will be quoted
are not formal errors but represent reasonable ranges for the values of
these parameters. An attempt to detect this line is currently being made
by the EURD detector (Bowyer et al 1995, Morales et al 1997, Bowyer,
Edelstein \& Lampton 1997) which is on board the orbiting Spanish
satellite MINISAT 01.

To determine the intensity of the line one must know the opacity of the
medium surrounding the sun for photons just beyond the Lyman limit. In
fact the sun is known to be immersed in a partially neutral hydrogen
cloud, which is the central part of what is
 called the Local Interstellar Medium (Cox \& Reynolds 1987). The best
determinations of the volume density n$_{\rm HI}\,\rm {of}\, HI$ near the sun have
been derived from $HST$ observations of the Lyman $\alpha$ absorption line
in the spectra of Procyon $(l=214^\circ,
 b=13^\circ, d=3.5 {\rm pc})$ (Linsky et al 1995) and of $\epsilon
 \;{\rm Ind}\; (l=336^\circ, b=-48^\circ,
d=3.46 {\rm pc})$ (Wood, Alexander \& Linsky 1996). 
They obtained for the lines
of sight to these two stars $n_{\rm HI}=0.1065\pm0.0028$ cm$^{-3}$ and
$0.094\pm0.022$ cm$^{-3}$ respectively. The two stars lie in rather
different directions, so it is comforting that they lead to the same value
of $n_{\rm HI}$, namely $0.1$ cm$^{-3}$. The corresponding mean free path $l$
for a photon at the Lyman limit $\sim{1\over 2}$ pc, which is substantially
less than the distances to the two stars. Accordingly the flux in the line
at the sun, which is $n_{\nu} l\over\tau$, will be $350^{+350}_{-117}$
cm$^{-2}$ sec$^{-1}$, since $n_\nu=4.16\times10^7$ cm$^{-3}$ and
$\tau=2\pm1\times 10^{23}$ sec. 

The wavelength $\lambda$ of the line for our updated value of $E_\gamma$
is
\begin{equation}
\lambda= 905 \pm 7\AA.
\end{equation}
The error quoted is
 the uncertainty in the central wavelength, not the linewidth, which, as
we shall see, is about $1 \AA$. Unfortunately our predicted wavelength
falls right inside the position of a much stronger nightglow emission
feature which stretches from about $900\AA$ to $911\AA$, and is due to
the recombination of $OII$ in the Earth's outer atmosphere (Chakrabarti
1984, Chakrabarti, Kimble \& Bowyer 1984). This emission feature was
detected by the EUV spectrometer on board the STP78-1 satellite which was
launched in
 1979. The minimum intensity of the detected feature is 15 Rayleighs,
which is $\sim5\times10^4$ times greater than our predicted flux for the
decay line. Nevertheless it may be possible to observe this line if
sufficient data are available (Bowyer 1997). 

Finally we consider the expected width of the decay line. This width is
due to the velocity dispersion $v$ of the neutrinos producing the line. For
a simple isotropic isothermal sphere model of the neutrino halo of our
Galaxy one would have $v=\sqrt{3\over 2}v_{rot}$, where the asymptotic
rotation velocity $v_{rot}$ of the Galaxy can be taken to be about 220 km.
sec$^{-1}$ (Binney \& Tremaine 1987). Thus $v\sim270$ km. sec$^{-1}$ and so
$\Delta\lambda<1\AA$, which is much less than the wavelength resolution of
EURD. However, recently Cowsik, Ratnam \& Bhattacharjee (1996) have claimed
to have constructed a self-consistent model of the dark matter halo of our
Galaxy which requires $v$ to lie between 600 and 900 km. sec$^{-1}$. This
claim has been challenged by Evans (1997), Gates, Kamionkowski \& Turner
(1997) and Bienaym\'e \& Pichon (1997), and Cowsik et al (1997) have
replied to the first two criticisms. 

We do not wish to enter into this controversy here, and merely note that,
if the decay line could be detected and its width measured, one would be
able to deduce directly the velocity dispersion of the neutrinos. In this
connexion it should be noted that in our strongly flattened model for the
neutrino halo (Sciama 1997b), referred to in section 2, the velocity
dispersion, and so the linewidth, would depend strongly on direction. One
could imagine measuring this anisotropic effect in a future mission with
adequate wavelength resolution, if the decay line could be disentangled
from the OI emission feature. In addition, if the neutrino halo itself has
little or no rotation, one might be able to observe the Doppler effect
associated with the sun's rotation in the Galaxy, which would shift the
central wavelength of the line by nearly $\mp1\AA$ in directions parallel
and antiparallel to the sun's motion. 

Acknowledgments

I am grateful to Marek Abramowicz, Stuart Bowyer, Walter Dehnen, Piero
Madau, Ron Reynolds and Ewa Szuszkiewicz for their crucial help,and to the
referee Don York for his valuable criticisms.  This work has been
supported by the Ministry of Universities and Scientific and Technological
Research.

\end{document}